\documentclass[aps,reprint, prb, longbibliography, superscriptaddress, showpacs,nobibnotes]{revtex4-1}
\usepackage{graphicx}
\usepackage{bm,amsmath,amssymb}
\usepackage{dcolumn}
\usepackage{natbib,xcolor}
\usepackage[version=4]{mhchem}
\usepackage{upgreek}

\begin{document}
\title{Note on the interpretation of magnetic diffraction in NdAlSi: helical or fan?}

\author{Takashi Kurumaji}
\affiliation{Division of Physics, Mathematics and Astronomy, California Institute of Technology, Pasadena, CA, 91125}

\date{\today}
\begin{abstract}
We revisit the magnetic structure analysis reported in Nat. Mater. \textbf{20}, 1650 (2021) \cite{gaudet2021weyl}, which concluded that a Weyl semimetal candidate NdAlSi hosts a helical magnetic order.
This conclusion was based on magnetic neutron diffraction peaks corresponding to modulation vectors $\vec{k}_{in}=(1/3,1/3,0)$ and $\vec{k}_{out}=(2/3,2/3,0)$, attributed to in-plane and out-of-plane components of the magnetic moments, respectively.
Upon careful reanalysis, we suggest that a fan-type magnetic structure--rather than the helix--provides a more consistent interpretation of these data.
Unlike a helical structure, fan structures do not exhibit handedness.
The distinction has significant implications for interpreting the electromagnetic responses in this material.
We believe that our suggestions motivate a re-examination of magnetic structures in NdAlSi and other tetragonal siblings, where the interplay between magnetism and topology is under active investigation.
\end{abstract}

\keywords{magnetism}
\maketitle
A helical spin structure is a noncollinear magnetic ordering where magnetic moments rotate around a propagation axis, completing a full 360$^{\circ}$ rotation \cite{herpin1970theorie}.
This structure possesses a defined handedness--i.e., a chiral sense of rotation--which underlies a variety of emergent phenomena, ranging from multiferroicity \cite{cheong2007multiferroics} to spintronic functionalities \cite{jiang2020electric,yokouchi2020emergent}, and has attracted attention for its link to odd-parity magnetism \cite{hellenes2023p,song2025electrical}.
In constrast, a fan-type structure \cite{nagamiya1962magnetization} involves magnetic moments tilting back and forth without completing a rotation, thus lacking handedness. 
These two distinct magnetic configurations are typically distinguished via neutron diffraction, where precise data analysis is essential for accurate identification.

\begin{figure}[ht]
	\includegraphics[width = \columnwidth]{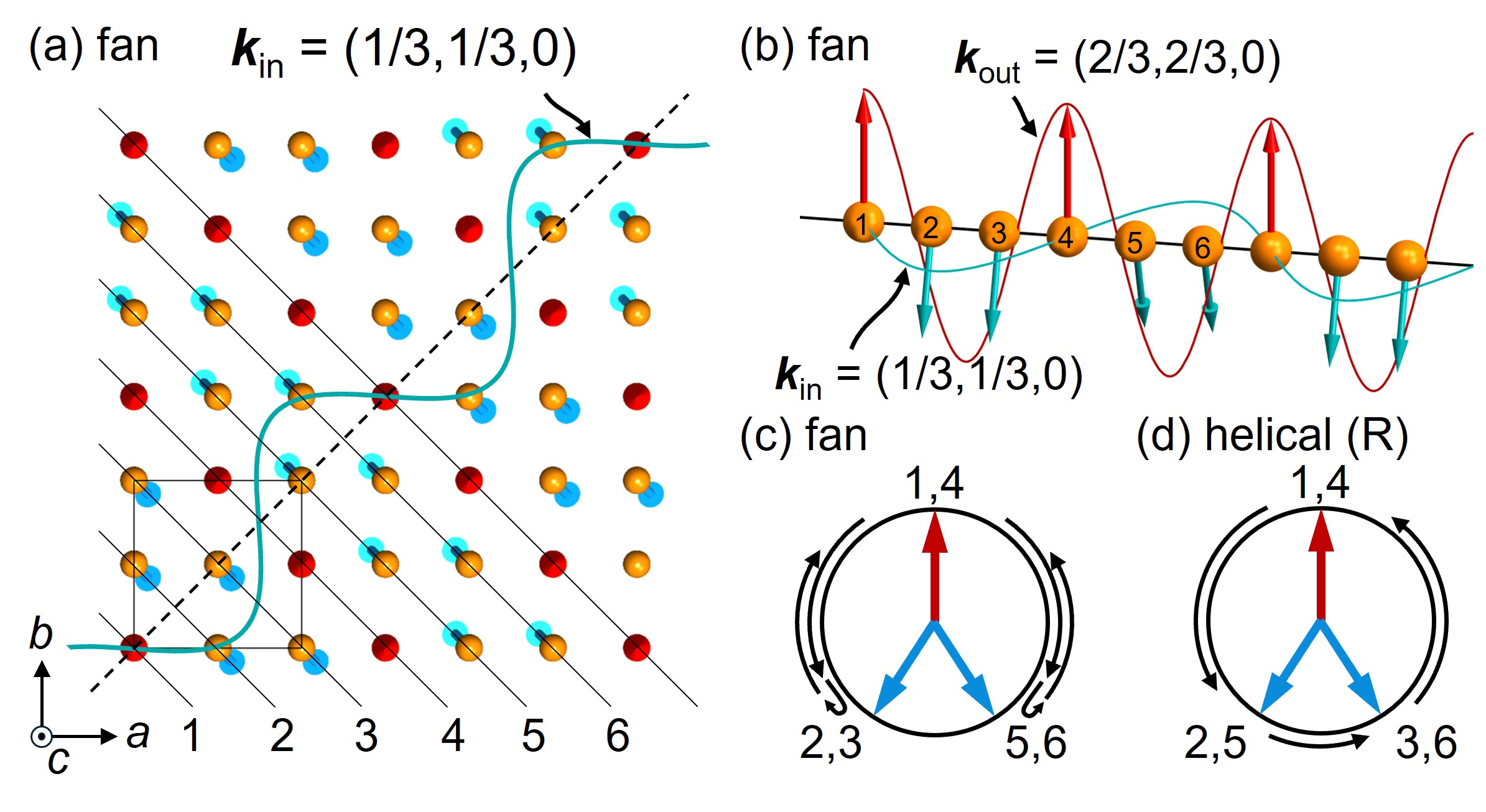}
	\caption{\label{fig1} (a) Top view of the Nd sublattice (orange sphere) in NdAlSi with the fan-type magnetic structure proposed in this note.
    Red arrows indicate magnetic moments pointing along the $+c$ axis, while cyan arrows are tilted from the $-c$ axis towards the $ab$ plane.
    Black square: the crystallographic unit cell; black lines: isophase surfaces perpendicular to $\vec{k}_{out/in}$, where the Nd sites at $\vec{r}_1$ ($\vec{r}_2$) is on the plane 1 (2).
    A cyan sine curve illustrates the in-plane component of the magnetic moments modulating with $\vec{k}_{in}$.
    (b) Schematic fan-type magnetic structure illustrated as an effective one-dimensional chain, projecting all Nd sites onto the $[110]$ line.
    Red (cyan) sinusoidal curve represents the modulating out-of-plane (in-plane) magnetic moments with $\vec{k}_{out}$ ($\vec{k}_{in}$).
    (c)-(d) Projection of magnetic moments on the $(110)$ plane for (c) the fan-type and (d) right-handed (R) helical structures.
    Numbers indicate the indices of the isophase planes defined in panel (a).
    }
\end{figure}

In this note, we revisit the interpretation of the magnetic structure in NdAlSi reported in a recent publication \cite{gaudet2021weyl}.
NdAlSi crystallizes in the polar LaPtSi-type structure with space group $I4_1md$.
Within a unit cell, there are two Nd sites located at $\vec{r}_1=(0,0,0)$ and $\vec{r}_2=(1/2,0,1/4)$ in the fractional coordinates, which are connected by a $d$-glide ($s_d=\{ m_{1\bar{1}0}|\frac{1}{4} \frac{1}{4} \frac{1}{4} \}$).
The key experimental finding in Ref.~\onlinecite{gaudet2021weyl} is the observation of magnetic Bragg scatterings at $\vec{Q}_{1/3}=(1/3,1/3,0)+\vec{G}$ and $\vec{Q}_{2/3}=(2/3,2/3,0)+\vec{G}$ in the reciprocal lattice unit ($\vec{G}$ is any reciprocal lattice vector corresponding to allowed nuclear Bragg peaks).
These scatterings were attributed to the in-plane and out-of-plane components of the magnetic moments, transversely modulating with the propagation vectors $\vec{k}_{in}=(1/3,1/3,0)$ and $\vec{k}_{out}=(2/3,2/3,0)$ respectively.
Based on the scattering intensity analysis, the authors interpreted these features as signatures of a helical structure, as illustrated in Fig. 3e of Ref.~\onlinecite{gaudet2021weyl}.

We revisit this interpretation and point out that a genuine helical structure requires a single modulation vector for both moment components. 
A general expression for a helical modulation is
\begin{equation}\label{heli}
    \vec{m}_{heli}(\vec{r})=(\vec{m}_{out}+i\vec{m}_{in})\exp (\vec{k}\cdot \vec{r}+\phi)+c.c.,
\end{equation}
where $\vec{m}_{out}$ and $\vec{m}_{in}$ are orthogonal and confined to the plane perpendicular to $\vec{k}$, and $\phi$ is the phase factor.
However, in Ref.~\onlinecite{gaudet2021weyl}, the in-plane and out-of-plane components exhibit different periodicities ($\vec{k}_{in}$ and $\vec{k}_{out}$), precluding a helical interpretation.

Instead, we propose a fan-type structure, as illustrated in Fig. \ref{fig1}(a).
The out-of-plane component forms an up-down-down modulation (consistent with $\vec{k}_{out}$), while the in-plane component modulates with $\vec{k}_{in}$, requiring six magnetic sublattices to complete one full period due to the presence of two Nd sites per unit cell.
The distinction becomes more apparent when represented in an effective one-dimensional magnetic chain (Fig. \ref{fig1}(b)).

This key difference becomes more evident when projecting the magnetic moments onto the $(110)$ plane (Figs. \ref{fig1}(c)-(d)).
In the fan type structure, the magnetic moments on plane 1 (as defined in Fig. \ref{fig1}(a)) point along the $+c$-axis.
As one moves across successive planes along the $[110]$ direction, the moments first rotate leftward on planes 2 and 3, return to the $+c$ direction on plane 4, and then rotate rightward on planes 5 and 6  (Fig. \ref{fig1}(c)).
This sequence of 0-left-left-0-right-right rotation (where, ``0'' denotes alignment along $+c$) repeats periodically.
Importantly, such a pattern does not trace a closed path in the spin space, and therefore lacks a chiral sense of rotation.
In contrast, the helical structure shown in Fig. 3e of Ref.~\onlinecite{gaudet2021weyl} is projected to Fig. \ref{fig1}(d), which exhibits unidirectional rotation, thereby defining the handedness.

The achiral nature of the crystallographic lattice preserves the degeneracy between right/left-handed helical configurations.
Application of the combined $\mathcal{T}s_d$ operation ($\mathcal{T}$: time reversal) yields the opposite chirality while maintaining the direction of the net ferromagnetic moment.
In contrast, the fan structure remains invariant under this operation, aside from a lattice translation.
These differences correspond to magnetic space groups $C2$ (chiral) for helical and $Cc'$ (achiral) for the fan structures, as determined using the k-SUBGROUPSMAG program on the Bilbao Crystallographic Server \cite{perez2015symmetry}.

To evaluate the compatibility between the experimental observations and candidate magnetic structures, we calculate the magnetic structure factor ($\vec{F}_M$) for each model.
$\vec{F}_M$ for the elastic neutron magnetic scattering with unpolarized neutrons is given by
\begin{equation}
    \vec{F}_M(\vec{Q})=\sum _j (\vec{S}_j-(\vec{S}_j\cdot \hat{\vec{Q}}) \hat{\vec{Q}})e^{i\vec{Q}\cdot \vec{r}_{Mj}},
\end{equation}
where $\vec{S}_j$ is the magnetic moment at $j$-th site in a magnetic unit cell, $\vec{Q}$ is the magnetic Bragg scattering vector (unit vector $\hat{\vec{Q}}$), and $\vec{r}_{Mj}$ is the position of the $j$-th magnetic site.
The scattering intensity is obtained by $I(\vec{Q})\propto |\vec{F}_M (\vec{Q})|^2$. 
Since the magnetic modulation is commensurate with the chemical lattice, the structural factor can be computed directly by summing over all magnetic sites within the magnetic unit cell.

The magnetic unit cell and the Nd site indexing for the helical and fan models are shown in Fig. 2.
We consider twelve inequivalent Nd sites per magnetic unit cell.
The Cartesian coordinates and magnetic moment vectors used in this calculation are summarized in Table~\ref{table}.
The ``wobbling'' modulation of the moment along the propagation vector--discussed in Ref.~\onlinecite{lygouras2024magnetic}--is not considered here\footnote{This longitudinal modulation is unlikely to affect our current analysis.
In the fan structure, the introduced modulation is $\vec{k}_{long}=(2/3,2/3,0)$, which contributes to scattering at $\vec{Q}_{2/3}=(2/3,2/3,\ell)$ with $\ell\neq 0$.
This modification imparts a cycloidal character, consistent with the crystal's polar nature.
Unlike helices, the cycloidal's handedness is fixed by the intrinsic lattice polarity.
}.
In both models, the magnetic unit cell is defined by the lattice vectors: $\vec{a}_{M1}=(2a,a,0)$, $\vec{a}_{M2}=(a,2a,0)$, and $\vec{a}_{M3}=(0,0,c)$, using Cartesian coordinates aligned with the tetragonal crystallographic axes: $\vec{a}_1=(a,0,0)$, $\vec{a}_2=(0,a,0)$, and $\vec{a}_3=(0,0,c)$, where $a$ and $c$ are the lattice constants.
The corresponding reciprocal lattice vectors are $\vec{b}_{M1}=\frac{2\pi}{3a}(2,-1,0)$, $\vec{b}_{M2}=\frac{2\pi}{3a}(-1,2,0)$, and $\vec{b}_{M3}=\frac{2\pi}{c}(0,0,1)$.
For simplicity, we consider the magnetic Bragg peaks located at $\vec{Q}_{1/3}=\vec{b}_{M1}+\vec{b}_{M2}$ and $\vec{Q}_{2/3}=2\vec{b}_{M1}+2\vec{b}_{M2}$, respectively, i.e., $\vec{G}=0$.

\begin{table*}
\caption{\label{table} Nd site index ($j$), magnetic site position ($\vec{r}_{Mj}$), magnetic moment vector $\vec{S}_j$ for the R-helical and fan models, and $\exp{(i\vec{Q}\cdot \vec{r}_{Mj})}$ at $\vec{Q}=\vec{Q}_{1/3}$ ($=\frac{2\pi}{3a}(1,1,0)$) and $\vec{Q}_{2/3}$ ($=\frac{2\pi}{3a}(2,2,0)$).
$S_x$, $S_z$, and $S_z'$ denote positive real numbers corresponding to components of the magnetic moment.
Since  all magnetic moments are transverse to the above scattering vectors in both models, $\vec{S}_{j}$ and $\vec{S}_j-(\vec{S}_j\cdot \hat{\vec{Q}})\hat{\vec{Q}}$ are identical.
The bottom rows presents the calculated $\vec{F}_M(\vec{Q})$ at each $\vec{Q}$ for both models.
}
\begin{tabular}{l*{7}{c}}
\hline
\hline
$j$ & $\vec{r}_{Mj}$ & $\vec{S}_j$ (Helical (R)) &$\vec{S}_j$ (Fan) &$\quad$& $\exp{(i\vec{Q}_{1/3}\cdot \vec{r}_{Mj})}$&$\quad$&$\exp{(i\vec{Q}_{2/3}\cdot \vec{r}_{Mj})}$ \\
\hline
\hline
$1$&$(0,0,0)$&$(0,0,S_z)$&$(0,0,S_z)$& &$1$& &$1$\\
$2$&$(a/2,a/2,c/2)$&$(-S_x,S_x,-S_z')$&$(S_x,-S_x,-S_z')$&&$e^{i2\pi/3}$&&$e^{i4\pi/3}$\\
$3$&$(a,a/2,3c/4)$&$(0,0,S_z)$&$(0,0,S_z)$&&$-1$&&$1$\\
$4$&$(a,a,0)$&$(S_x,-S_x,-S_z')$&$(-S_x,S_x,-S_z')$&&$e^{i4\pi/3}$&&$e^{i2\pi/3}$\\
$5$&$(3a/2,a,c/4)$&$(-S_x,S_x,-S_z')$&$(-S_x,S_x,-S_z')$&&$-e^{i2\pi/3}$&&$e^{i4\pi/3}$\\
$6$&$(a,3a/2,3a/4)$&$(-S_x,S_x,-S_z')$&$(-S_x,S_x,-S_z')$&&$-e^{i2\pi/3}$&&$e^{i4\pi/3}$\\
$7$&$(3a/2,3a/2,c/2)$&$(0,0,S_z)$&$(0,0,S_z)$&&$1$&&$1$\\
$8$&$(2a,3a/2,3c/4)$&$(S_x,-S_x,-S_z')$&$(S_x,-S_x,-S_z')$&&$-e^{i4\pi/3}$&&$e^{i2\pi/3}$\\
$9$&$(3a/2,2a,c/4)$&$(S_x,-S_x,-S_z')$&$(S_x,-S_x,-S_z')$&&$-e^{i4\pi/3}$&&$e^{i2\pi/3}$\\
$10$&$(2a,2a,0)$&$(-S_x,S_x,-S_z')$&$(S_x,-S_x,-S_z')$&&$e^{i2\pi/3}$&&$e^{i4\pi/3}$\\
$11$&$(5a/2,2a,c/4)$&$(0,0,S_z)$&$(0,0,S_z)$&&$-1$&&$1$\\
$12$&$(5a/2,5a/2,c/2)$&$(S_x,-S_x,-S_z')$&$(-S_x,S_x,-S_z')$&&$e^{i4\pi/3}$&&$e^{i2\pi/3}$\\
\hline
\hline
\multicolumn{7}{c}{$\vec{F}_M(\vec{Q})=\sum _{j=1}^{12} (\vec{S}_j-(\vec{S}_j\cdot \hat{\vec{Q}}) \hat{\vec{Q}})e^{i\vec{Q}\cdot \vec{r}_{Mj}}$}\\
\hline
&Observation\footnote{see Supplementary Fig. S3 in Ref.~\onlinecite{gaudet2021weyl}}&Helical (R)&Fan&&&&\\
$\vec{Q}_{1/3}$&$\checkmark$ (in-plane moment)& $(0,0,0)$&$(i4\sqrt{3}S_x,-i4\sqrt{3}S_x,0)$&\\
$\vec{Q}_{2/3}$&$\checkmark$ (out-of-plane moment)&$(i4\sqrt{3}S_x,-i4\sqrt{3}S_x,4S_z+4S_z')$ &$(0,0,4S_z+4S_z')$ &\\
\hline
\hline
\end{tabular}
\end{table*}

Our calculations of $\vec{F}_M$ (Table~\ref{table}) show that the helical model yields a finite intensity at $\vec{Q}_{2/3}$ but vanishing intensity at $\vec{Q}_{1/3}$, contradicting experimental results.
This is reasonable because the helical model is described by the in-plane ($S_x$) and out-of-plane ($S_z$ and $S_z'$) magnetic moments with the common modulation $\vec{k}_{out}$ as in Eq. (\ref{heli}).
In contrast, the fan model predicts finite intensities at both $\vec{Q}_{1/3}$ and $\vec{Q}_{2/3}$, with the former arising solely from the $S_x$ and the latter from $S_z$ and $S_z'$.
This is in good agreement with the decomposed modulations shown in Fig. \ref{fig1}(b) and the observation in Ref.~\onlinecite{gaudet2021weyl}.

Upon careful reanalysis of the experimental data, we propose that the magnetic structure of NdAlSi is more appropriately described by a fan-type structure rather than a helical configuration.
As the emergence of a helical magnetic structure is the central claim of the original publication--highlighted in both the title of the article and News \& Views \cite{hirschberger2021weyl}--any ambiguity in its interpretation is of critical importance for the understanding of electromagnetic responses in this material class.
We therefore respectfully call for a careful re-examination of the magnetic structure in NdAlSi, as well as in related isostructural compounds \cite{yao2023large,yang2023stripe}.

\begin{figure}[ht]
	\includegraphics[width = 1\columnwidth]{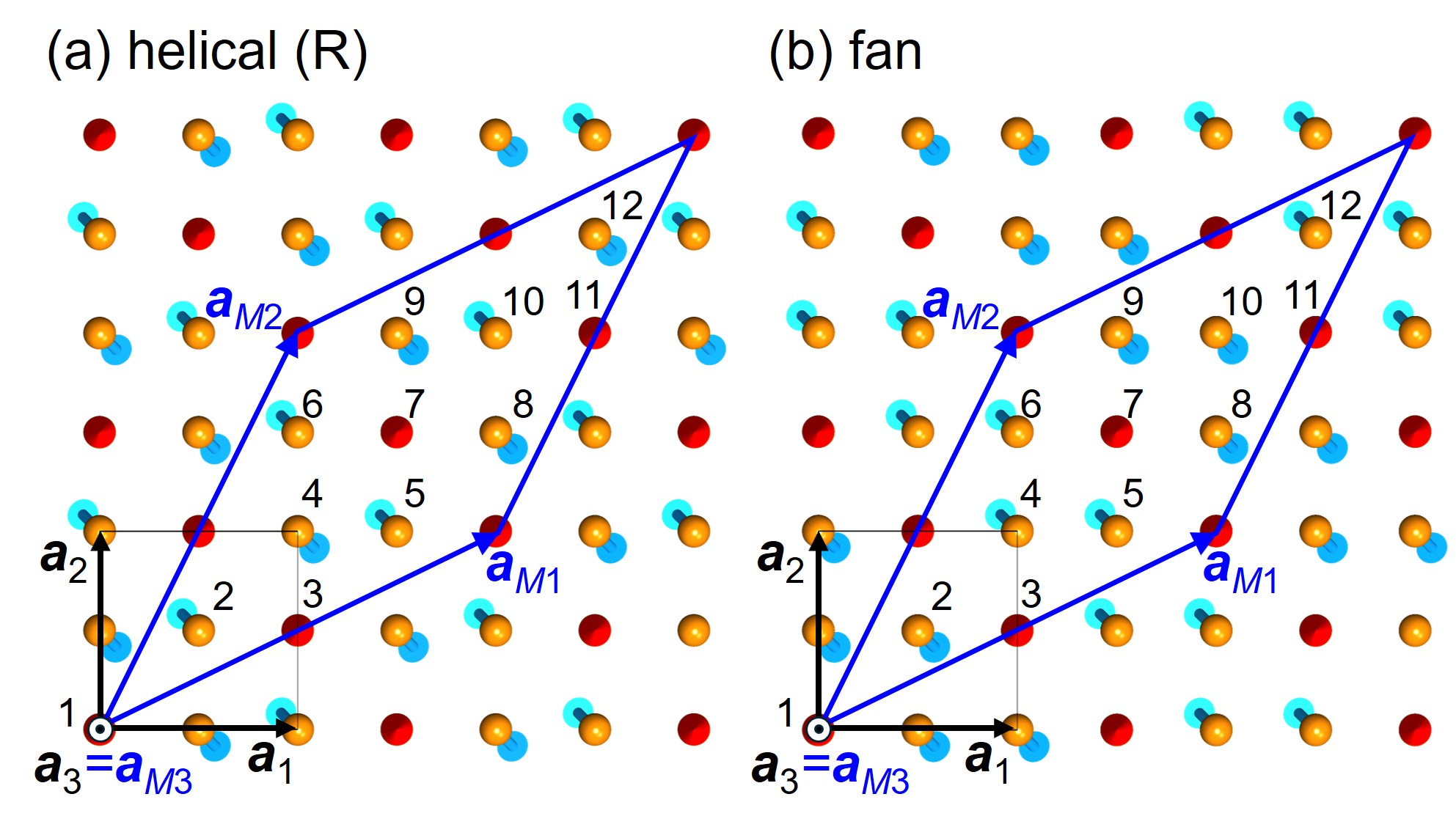}
	\caption{\label{fig2} Lattice vectors ($\vec{a}_i$, $\vec{a}_{Mi}$, $i=1,2,3$) for the crystallographic (black square) and magnetic unit cells (blue rhombus) in (a) R-handed helical model and (b) fan-type model.
    Numbers denote the indices of the magnetic sites at $\vec{r}_{Mj}$.}
\end{figure}

\begin{acknowledgements}
We are grateful to Dr. J. Gaudet for helpful suggestions, and we sincerely thank the anonymous author of a scientific blog for valuable insights and encouragement, which motivate us to prepare this manuscript.
\end{acknowledgements}

%


\begin{thebibliography}{14}%
\makeatletter
\providecommand \@ifxundefined [1]{%
 \@ifx{#1\undefined}
}%
\providecommand \@ifnum [1]{%
 \ifnum #1\expandafter \@firstoftwo
 \else \expandafter \@secondoftwo
 \fi
}%
\providecommand \@ifx [1]{%
 \ifx #1\expandafter \@firstoftwo
 \else \expandafter \@secondoftwo
 \fi
}%
\providecommand \natexlab [1]{#1}%
\providecommand \enquote  [1]{``#1''}%
\providecommand \bibnamefont  [1]{#1}%
\providecommand \bibfnamefont [1]{#1}%
\providecommand \citenamefont [1]{#1}%
\providecommand \href@noop [0]{\@secondoftwo}%
\providecommand \href [0]{\begingroup \@sanitize@url \@href}%
\providecommand \@href[1]{\@@startlink{#1}\@@href}%
\providecommand \@@href[1]{\endgroup#1\@@endlink}%
\providecommand \@sanitize@url [0]{\catcode `\\12\catcode `\$12\catcode `\&12\catcode `\#12\catcode `\^12\catcode `\_12\catcode `\%12\relax}%
\providecommand \@@startlink[1]{}%
\providecommand \@@endlink[0]{}%
\providecommand \url  [0]{\begingroup\@sanitize@url \@url }%
\providecommand \@url [1]{\endgroup\@href {#1}{\urlprefix }}%
\providecommand \urlprefix  [0]{URL }%
\providecommand \Eprint [0]{\href }%
\providecommand \doibase [0]{http://dx.doi.org/}%
\providecommand \selectlanguage [0]{\@gobble}%
\providecommand \bibinfo  [0]{\@secondoftwo}%
\providecommand \bibfield  [0]{\@secondoftwo}%
\providecommand \translation [1]{[#1]}%
\providecommand \BibitemOpen [0]{}%
\providecommand \bibitemStop [0]{}%
\providecommand \bibitemNoStop [0]{.\EOS\space}%
\providecommand \EOS [0]{\spacefactor3000\relax}%
\providecommand \BibitemShut  [1]{\csname bibitem#1\endcsname}%
\let\auto@bib@innerbib\@empty
\bibitem [{\citenamefont {Gaudet}\ \emph {et~al.}(2021)\citenamefont {Gaudet}, \citenamefont {Yang}, \citenamefont {Baidya}, \citenamefont {Lu}, \citenamefont {Xu}, \citenamefont {Zhao}, \citenamefont {Rodriguez-Rivera}, \citenamefont {Hoffmann}, \citenamefont {Graf}, \citenamefont {Torchinsky} \emph {et~al.}}]{gaudet2021weyl}%
  \BibitemOpen
  \bibfield  {author} {\bibinfo {author} {\bibfnamefont {J.}~\bibnamefont {Gaudet}}, \bibinfo {author} {\bibfnamefont {H.-Y.}\ \bibnamefont {Yang}}, \bibinfo {author} {\bibfnamefont {S.}~\bibnamefont {Baidya}}, \bibinfo {author} {\bibfnamefont {B.}~\bibnamefont {Lu}}, \bibinfo {author} {\bibfnamefont {G.}~\bibnamefont {Xu}}, \bibinfo {author} {\bibfnamefont {Y.}~\bibnamefont {Zhao}}, \bibinfo {author} {\bibfnamefont {J.~A.}\ \bibnamefont {Rodriguez-Rivera}}, \bibinfo {author} {\bibfnamefont {C.~M.}\ \bibnamefont {Hoffmann}}, \bibinfo {author} {\bibfnamefont {D.~E.}\ \bibnamefont {Graf}}, \bibinfo {author} {\bibfnamefont {D.~H.}\ \bibnamefont {Torchinsky}},  \emph {et~al.},\ }\bibfield  {title} {\enquote {\bibinfo {title} {Weyl-mediated helical magnetism in {N}d{A}l{S}i},}\ }\href@noop {} {\bibfield  {journal} {\bibinfo  {journal} {Nat. Mater.}\ }\textbf {\bibinfo {volume} {20}},\ \bibinfo {pages} {1650} (\bibinfo {year} {2021})}\BibitemShut {NoStop}%
\bibitem [{\citenamefont {Herpin}\ and\ \citenamefont {Mattis}(1970)}]{herpin1970theorie}%
  \BibitemOpen
  \bibfield  {author} {\bibinfo {author} {\bibfnamefont {A.}~\bibnamefont {Herpin}}\ and\ \bibinfo {author} {\bibfnamefont {D.~C.}\ \bibnamefont {Mattis}},\ }\href@noop {} {\enquote {\bibinfo {title} {Th{\'e}orie du magn{\'e}tisme},}\ } (\bibinfo {year} {1970})\BibitemShut {NoStop}%
\bibitem [{\citenamefont {Cheong}\ and\ \citenamefont {Mostovoy}(2007)}]{cheong2007multiferroics}%
  \BibitemOpen
  \bibfield  {author} {\bibinfo {author} {\bibfnamefont {S.-W.}\ \bibnamefont {Cheong}}\ and\ \bibinfo {author} {\bibfnamefont {M.}~\bibnamefont {Mostovoy}},\ }\bibfield  {title} {\enquote {\bibinfo {title} {Multiferroics: a magnetic twist for ferroelectricity},}\ }\href@noop {} {\bibfield  {journal} {\bibinfo  {journal} {Nat. Mater.}\ }\textbf {\bibinfo {volume} {6}},\ \bibinfo {pages} {13} (\bibinfo {year} {2007})}\BibitemShut {NoStop}%
\bibitem [{\citenamefont {Jiang}\ \emph {et~al.}(2020)\citenamefont {Jiang}, \citenamefont {Nii}, \citenamefont {Arisawa}, \citenamefont {Saitoh},\ and\ \citenamefont {Onose}}]{jiang2020electric}%
  \BibitemOpen
  \bibfield  {author} {\bibinfo {author} {\bibfnamefont {N.}~\bibnamefont {Jiang}}, \bibinfo {author} {\bibfnamefont {Y.}~\bibnamefont {Nii}}, \bibinfo {author} {\bibfnamefont {H.}~\bibnamefont {Arisawa}}, \bibinfo {author} {\bibfnamefont {E.}~\bibnamefont {Saitoh}}, \ and\ \bibinfo {author} {\bibfnamefont {Y.}~\bibnamefont {Onose}},\ }\bibfield  {title} {\enquote {\bibinfo {title} {Electric current control of spin helicity in an itinerant helimagnet},}\ }\href@noop {} {\bibfield  {journal} {\bibinfo  {journal} {Nat. Commun.}\ }\textbf {\bibinfo {volume} {11}},\ \bibinfo {pages} {1601} (\bibinfo {year} {2020})}\BibitemShut {NoStop}%
\bibitem [{\citenamefont {Yokouchi}\ \emph {et~al.}(2020)\citenamefont {Yokouchi}, \citenamefont {Kagawa}, \citenamefont {Hirschberger}, \citenamefont {Otani}, \citenamefont {Nagaosa},\ and\ \citenamefont {Tokura}}]{yokouchi2020emergent}%
  \BibitemOpen
  \bibfield  {author} {\bibinfo {author} {\bibfnamefont {T.}~\bibnamefont {Yokouchi}}, \bibinfo {author} {\bibfnamefont {F.}~\bibnamefont {Kagawa}}, \bibinfo {author} {\bibfnamefont {M.}~\bibnamefont {Hirschberger}}, \bibinfo {author} {\bibfnamefont {Y.}~\bibnamefont {Otani}}, \bibinfo {author} {\bibfnamefont {N.}~\bibnamefont {Nagaosa}}, \ and\ \bibinfo {author} {\bibfnamefont {Y.}~\bibnamefont {Tokura}},\ }\bibfield  {title} {\enquote {\bibinfo {title} {Emergent electromagnetic induction in a helical-spin magnet},}\ }\href@noop {} {\bibfield  {journal} {\bibinfo  {journal} {Nature}\ }\textbf {\bibinfo {volume} {586}},\ \bibinfo {pages} {232} (\bibinfo {year} {2020})}\BibitemShut {NoStop}%
\bibitem [{\citenamefont {Hellenes}\ \emph {et~al.}(2023)\citenamefont {Hellenes}, \citenamefont {Jungwirth}, \citenamefont {Jaeschke-Ubiergo}, \citenamefont {Chakraborty}, \citenamefont {Sinova},\ and\ \citenamefont {{\v{S}}mejkal}}]{hellenes2023p}%
  \BibitemOpen
  \bibfield  {author} {\bibinfo {author} {\bibfnamefont {A.~B.}\ \bibnamefont {Hellenes}}, \bibinfo {author} {\bibfnamefont {T.}~\bibnamefont {Jungwirth}}, \bibinfo {author} {\bibfnamefont {R.}~\bibnamefont {Jaeschke-Ubiergo}}, \bibinfo {author} {\bibfnamefont {A.}~\bibnamefont {Chakraborty}}, \bibinfo {author} {\bibfnamefont {J.}~\bibnamefont {Sinova}}, \ and\ \bibinfo {author} {\bibfnamefont {L.}~\bibnamefont {{\v{S}}mejkal}},\ }\bibfield  {title} {\enquote {\bibinfo {title} {P-wave magnets},}\ }\href@noop {} {\bibfield  {journal} {\bibinfo  {journal} {arXiv preprint arXiv:2309.01607}\ } (\bibinfo {year} {2023})}\BibitemShut {NoStop}%
\bibitem [{\citenamefont {Song}\ \emph {et~al.}(2025)\citenamefont {Song}, \citenamefont {Stavri{\'c}}, \citenamefont {Barone}, \citenamefont {Droghetti}, \citenamefont {Antonenko}, \citenamefont {Venderbos}, \citenamefont {Occhialini}, \citenamefont {Ilyas}, \citenamefont {Erge{\c{c}}en}, \citenamefont {Gedik} \emph {et~al.}}]{song2025electrical}%
  \BibitemOpen
  \bibfield  {author} {\bibinfo {author} {\bibfnamefont {Q.}~\bibnamefont {Song}}, \bibinfo {author} {\bibfnamefont {S.}~\bibnamefont {Stavri{\'c}}}, \bibinfo {author} {\bibfnamefont {P.}~\bibnamefont {Barone}}, \bibinfo {author} {\bibfnamefont {A.}~\bibnamefont {Droghetti}}, \bibinfo {author} {\bibfnamefont {D.~S.}\ \bibnamefont {Antonenko}}, \bibinfo {author} {\bibfnamefont {J.~W.~F.}\ \bibnamefont {Venderbos}}, \bibinfo {author} {\bibfnamefont {C.~A.}\ \bibnamefont {Occhialini}}, \bibinfo {author} {\bibfnamefont {B.}~\bibnamefont {Ilyas}}, \bibinfo {author} {\bibfnamefont {E.}~\bibnamefont {Erge{\c{c}}en}}, \bibinfo {author} {\bibfnamefont {N.}~\bibnamefont {Gedik}},  \emph {et~al.},\ }\bibfield  {title} {\enquote {\bibinfo {title} {Electrical switching of a p-wave magnet},}\ }\href@noop {} {\bibfield  {journal} {\bibinfo  {journal} {Nature}\ } (\bibinfo {year} {2025})}\BibitemShut {NoStop}%
\bibitem [{\citenamefont {Nagamiya}\ \emph {et~al.}(1962)\citenamefont {Nagamiya}, \citenamefont {Nagata},\ and\ \citenamefont {Kitano}}]{nagamiya1962magnetization}%
  \BibitemOpen
  \bibfield  {author} {\bibinfo {author} {\bibfnamefont {T.}~\bibnamefont {Nagamiya}}, \bibinfo {author} {\bibfnamefont {K.}~\bibnamefont {Nagata}}, \ and\ \bibinfo {author} {\bibfnamefont {Y.}~\bibnamefont {Kitano}},\ }\bibfield  {title} {\enquote {\bibinfo {title} {Magnetization process of a screw spin system},}\ }\href@noop {} {\bibfield  {journal} {\bibinfo  {journal} {Prog. Theor. Phys.}\ }\textbf {\bibinfo {volume} {27}},\ \bibinfo {pages} {1253} (\bibinfo {year} {1962})}\BibitemShut {NoStop}%
\bibitem [{\citenamefont {Perez-Mato}\ \emph {et~al.}(2015)\citenamefont {Perez-Mato}, \citenamefont {Gallego}, \citenamefont {Tasci}, \citenamefont {Elcoro}, \citenamefont {de~la Flor},\ and\ \citenamefont {Aroyo}}]{perez2015symmetry}%
  \BibitemOpen
  \bibfield  {author} {\bibinfo {author} {\bibfnamefont {J.~M.}\ \bibnamefont {Perez-Mato}}, \bibinfo {author} {\bibfnamefont {S.~V.}\ \bibnamefont {Gallego}}, \bibinfo {author} {\bibfnamefont {E.~S.}\ \bibnamefont {Tasci}}, \bibinfo {author} {\bibfnamefont {L.}~\bibnamefont {Elcoro}}, \bibinfo {author} {\bibfnamefont {G.}~\bibnamefont {de~la Flor}}, \ and\ \bibinfo {author} {\bibfnamefont {M.~I.}\ \bibnamefont {Aroyo}},\ }\bibfield  {title} {\enquote {\bibinfo {title} {Symmetry-based computational tools for magnetic crystallography},}\ }\href@noop {} {\bibfield  {journal} {\bibinfo  {journal} {Annu. Rev. Mater. Res.}\ }\textbf {\bibinfo {volume} {45}},\ \bibinfo {pages} {217} (\bibinfo {year} {2015})}\BibitemShut {NoStop}%
\bibitem [{\citenamefont {Lygouras}\ \emph {et~al.}(2024)\citenamefont {Lygouras}, \citenamefont {Yang}, \citenamefont {Yao}, \citenamefont {Gaudet}, \citenamefont {Hao}, \citenamefont {Cao}, \citenamefont {Rodriguez-Rivera}, \citenamefont {Podlesnyak}, \citenamefont {Bl{\"u}gel}, \citenamefont {Nikoli{\'c}} \emph {et~al.}}]{lygouras2024magnetic}%
  \BibitemOpen
  \bibfield  {author} {\bibinfo {author} {\bibfnamefont {C.~J.}\ \bibnamefont {Lygouras}}, \bibinfo {author} {\bibfnamefont {H.-Y.}\ \bibnamefont {Yang}}, \bibinfo {author} {\bibfnamefont {X.}~\bibnamefont {Yao}}, \bibinfo {author} {\bibfnamefont {J.}~\bibnamefont {Gaudet}}, \bibinfo {author} {\bibfnamefont {Y.}~\bibnamefont {Hao}}, \bibinfo {author} {\bibfnamefont {H.}~\bibnamefont {Cao}}, \bibinfo {author} {\bibfnamefont {J.~A.}\ \bibnamefont {Rodriguez-Rivera}}, \bibinfo {author} {\bibfnamefont {A.}~\bibnamefont {Podlesnyak}}, \bibinfo {author} {\bibfnamefont {S.}~\bibnamefont {Bl{\"u}gel}}, \bibinfo {author} {\bibfnamefont {P.}~\bibnamefont {Nikoli{\'c}}},  \emph {et~al.},\ }\bibfield  {title} {\enquote {\bibinfo {title} {Magnetic excitations and interactions in the {W}eyl ferrimagnet {N}d{A}l{S}i},}\ }\href@noop {} {\bibfield  {journal} {\bibinfo  {journal} {arXiv preprint arXiv:2412.20743}\ } (\bibinfo {year} {2024})}\BibitemShut {NoStop}%
\bibitem [{Note1()}]{Note1}%
  \BibitemOpen
  \bibinfo {note} {This longitudinal modulation is unlikely to affect our current analysis. In the fan structure, the introduced modulation is $\protect \vec {k}_{long}=(2/3,2/3,0)$, which contributes to scattering at $\protect \vec {Q}_{2/3}=(2/3,2/3,\ell )$ with $\ell \protect \neq 0$. This modification imparts a cycloidal character, consistent with the crystal's polar nature. Unlike helices, the cycloidal's handedness is fixed by the intrinsic lattice polarity.}\BibitemShut {Stop}%
\bibitem [{\citenamefont {Hirschberger}\ and\ \citenamefont {Tokura}(2021)}]{hirschberger2021weyl}%
  \BibitemOpen
  \bibfield  {author} {\bibinfo {author} {\bibfnamefont {M.}~\bibnamefont {Hirschberger}}\ and\ \bibinfo {author} {\bibfnamefont {Y.}~\bibnamefont {Tokura}},\ }\bibfield  {title} {\enquote {\bibinfo {title} {Weyl fermions promote collective magnetism},}\ }\href@noop {} {\bibfield  {journal} {\bibinfo  {journal} {Nat. Mater.}\ }\textbf {\bibinfo {volume} {20}},\ \bibinfo {pages} {1592} (\bibinfo {year} {2021})}\BibitemShut {NoStop}%
\bibitem [{\citenamefont {Yao}\ \emph {et~al.}(2023)\citenamefont {Yao}, \citenamefont {Gaudet}, \citenamefont {Verma}, \citenamefont {Graf}, \citenamefont {Yang}, \citenamefont {Bahrami}, \citenamefont {Zhang}, \citenamefont {Aczel}, \citenamefont {Subedi}, \citenamefont {Torchinsky} \emph {et~al.}}]{yao2023large}%
  \BibitemOpen
  \bibfield  {author} {\bibinfo {author} {\bibfnamefont {X.}~\bibnamefont {Yao}}, \bibinfo {author} {\bibfnamefont {J.}~\bibnamefont {Gaudet}}, \bibinfo {author} {\bibfnamefont {R.}~\bibnamefont {Verma}}, \bibinfo {author} {\bibfnamefont {D.~E.}\ \bibnamefont {Graf}}, \bibinfo {author} {\bibfnamefont {H.-Y.}\ \bibnamefont {Yang}}, \bibinfo {author} {\bibfnamefont {F.}~\bibnamefont {Bahrami}}, \bibinfo {author} {\bibfnamefont {R.}~\bibnamefont {Zhang}}, \bibinfo {author} {\bibfnamefont {A.~A.}\ \bibnamefont {Aczel}}, \bibinfo {author} {\bibfnamefont {S.}~\bibnamefont {Subedi}}, \bibinfo {author} {\bibfnamefont {D.~H.}\ \bibnamefont {Torchinsky}},  \emph {et~al.},\ }\bibfield  {title} {\enquote {\bibinfo {title} {Large topological {H}all effect and spiral magnetic order in the {W}eyl semimetal {S}m{A}l{S}i},}\ }\href@noop {} {\bibfield  {journal} {\bibinfo  {journal} {Phys. Rev. X}\ }\textbf {\bibinfo {volume} {13}},\ \bibinfo {pages} {011035} (\bibinfo {year} {2023})}\BibitemShut {NoStop}%
\bibitem [{\citenamefont {Yang}\ \emph {et~al.}(2023)\citenamefont {Yang}, \citenamefont {Gaudet}, \citenamefont {Verma}, \citenamefont {Baidya}, \citenamefont {Bahrami}, \citenamefont {Yao}, \citenamefont {Huang}, \citenamefont {DeBeer-Schmitt}, \citenamefont {Aczel}, \citenamefont {Xu} \emph {et~al.}}]{yang2023stripe}%
  \BibitemOpen
  \bibfield  {author} {\bibinfo {author} {\bibfnamefont {H.-Y.}\ \bibnamefont {Yang}}, \bibinfo {author} {\bibfnamefont {J.}~\bibnamefont {Gaudet}}, \bibinfo {author} {\bibfnamefont {R.}~\bibnamefont {Verma}}, \bibinfo {author} {\bibfnamefont {S.}~\bibnamefont {Baidya}}, \bibinfo {author} {\bibfnamefont {F.}~\bibnamefont {Bahrami}}, \bibinfo {author} {\bibfnamefont {X.}~\bibnamefont {Yao}}, \bibinfo {author} {\bibfnamefont {C.-Y.}\ \bibnamefont {Huang}}, \bibinfo {author} {\bibfnamefont {L.}~\bibnamefont {DeBeer-Schmitt}}, \bibinfo {author} {\bibfnamefont {A.~A.}\ \bibnamefont {Aczel}}, \bibinfo {author} {\bibfnamefont {G.}~\bibnamefont {Xu}},  \emph {et~al.},\ }\bibfield  {title} {\enquote {\bibinfo {title} {Stripe helical magnetism and two regimes of anomalous {H}all effect in {N}d{A}l{G}e},}\ }\href@noop {} {\bibfield  {journal} {\bibinfo  {journal} {Phys. Rev. Mater.}\ }\textbf {\bibinfo {volume} {7}},\ \bibinfo {pages} {034202} (\bibinfo {year} {2023})}\BibitemShut {NoStop}%
\end{thebibliography}

\end{document}